# Accelerated Discovery of Molten Salt Corrosion-resistant Alloy by High-throughput Experimental and Modeling Methods Coupled to Data Analytics

Yafei Wang[1*], Bonita Goh[1], Phalgun Nelaturu[1], Thien Duong[3], Najlaa Hassan[1], Raphaelle David[1], Michael Moorehead[1], Santanu Chaudhuri[3], Adam Creuziger[4], Jason Hattrick-Simpers[4], Dan J. Thoma[1,2], Kumar Sridharan[1,2], Adrien Couet[1,2]

[1]Department of Engineering Physics, University of Wisconsin-Madison, WI 53706, USA

[2]Department of Materials Science and Engineering, University of Wisconsin-Madison, WI 53706, USA

[3] Applied Materials Division, Argonne National Laboratory, Lemont, IL 60607, USA

[4] National Institute of Standard and Technology, Gaithersburg, MD 20899, USA

* wang749@wisc.edu

## Abstract

Insufficient availability of molten salt corrosion-resistant alloys severely limits the fruition of a variety of promising molten salt technologies that could otherwise have significant societal impacts. To accelerate alloy development for molten salt applications and develop fundamental understanding of corrosion in these environments, here we present an integrated approach using a set of high-throughput alloy synthesis, corrosion testing, and modeling coupled with automated characterization and machine learning. By using this approach, a broad range of Cr-Fe-Mn-Ni alloys were evaluated for their corrosion resistances in molten salt simultaneously demonstrating that corrosion-resistant alloy development can be accelerated by thousands of times. Based on the obtained results, we unveiled a sacrificial mechanism in the corrosion of Cr-Fe-Mn-Ni alloys in molten salts which can be applied to protect the less unstable elements in the alloy from being depleted, and provided new insights on the design of high-temperature molten salt corrosion-resistant alloys.

## Introduction

Molten salts have many attractive properties including: (i) relatively low melting point and high boiling point, (ii) high thermal conductivity, (iii) large heat capacity, and (iv) excellent compositional stability. These properties make them leading candidates for high-temperature applications such as coolant and fuel-solvent for nuclear reactors [1], extraction media for pyroprocessing of spent nuclear fuel [2], thermal energy storage and heat transfer fluid of concentrated solar power [3], and battery electrolytes [4]. One major concern in the deployment of these applications is material's corrosion in molten salts. Indeed, the protective oxide layer relied upon for the corrosion resistance in most aqueous solutions or oxidative environments is rendered unstable in high-temperature molten salts, especially molten chloride and fluoride salts [5]. Because of the strong electronegativity of chlorine and fluorine anions, these molten salts typically tend to destabilize the protective oxide layers, resulting in dissolution of the least thermodynamically stable element from the alloy into the salt. For example, Cr, one of the main constituents for all high-temperature (>700 °C) ASME code certified alloys, forms relatively stable metal halides and is particularly susceptible to dissolution in salts, such that all these alloys experience unacceptable corrosion rates in molten salts. Therefore, identifying corrosion resistant alloys compatible with high-temperature molten salt environments is crucial for the large-scale deployment of molten salt technologies.

Currently, the most promising alloy for high-temperature molten salt applications is Hastelloy-N, a Ni-based alloy, containing 16Mo-7Cr-Fe and other minor alloying elements [6]. However, this alloy has limited high-temperature creep resistance and its down-selection was based on the testing of nine alloys from the INOR series with varying Mo, Cr, Fe, Ti, Al, Nb and W contents [7]. One can legitimately question if no other alloys within this alloy system, or within other alloy systems, could exhibit higher corrosion resistance. However, current state-of-the-art alloy processing and testing in molten salt environments are not compatible with such an exploratory search and the immediate needs for molten salt technology deployment. For instance, arc melting [8], field assisted sintering or casting [9] can only melt one ingot composition at a time, and the manufactured materials have to be annealed, cut, and further prepared for testing. In addition, only one alloy can be tested at a time for its corrosion behavior in a given test cell, and moreover these corrosion experiments usually have to be performed continuously for up to a few thousand hours [10,11]. This paradigm results in an extremely lengthy processing/testing/characterization pathway if one wishes to explore a large compositional map for an alloy system. Consequently, the material development processes and screening for high-temperature molten salt technologies have lagged behind.

To accelerate the search for corrosion resistant alloys in molten salt and better understand the corrosion mechanism in molten salt across a wide range of conditions, high-throughput (HTP) experimental and modeling techniques have been developed and integrated in the present study as presented in Figure 1. First, an HTP alloy synthesis technique, namely in-situ alloying using additive manufacturing, was employed to print bulk alloys of different selected compositions. The laser engineered net shaping (LENS) process was adopted [12], such that any alloy composition can be manufactured in a few minutes. HTP corrosion tests were also designed by melting molten salt pills on top of each printed bulk alloy at 500 °C. The corrosion behavior of each printed alloy was assessed by a series of automated material characterization techniques such as glow-discharge optical emission spectroscopy (GDOES) and inductively coupled plasma mass spectrometry (ICP-MS). In addition, HTP first-principle density functional theory (DFT) calculations of surface energy and work function were coupled with a CALculation of PHAse Diagram (CALPHAD) method to rank the corrosion resistance of different printed alloys. Finally, features extraction and cross-validation were performed by machine learning (ML) based methods to determine the dependencies of corrosion resistance on alloys' physical parameters. This HTP platform has been experimented for the first time to exploit the correlation between the Mn and Fe dissolutions in Cr-Fe-Mn-Ni alloys and the alloys' corrosion resistance. A unique high temperature sacrificial corrosion mechanism in Cr-Fe-Mn-Ni alloys exposed to molten salt has been unveiled and discussed. The present study paves the way for accelerating the discovery of corrosion resistant alloys in molten salts, and serves as an example of alloy design for extreme environments using high-throughput and/or automated methods coupled to data analytics.

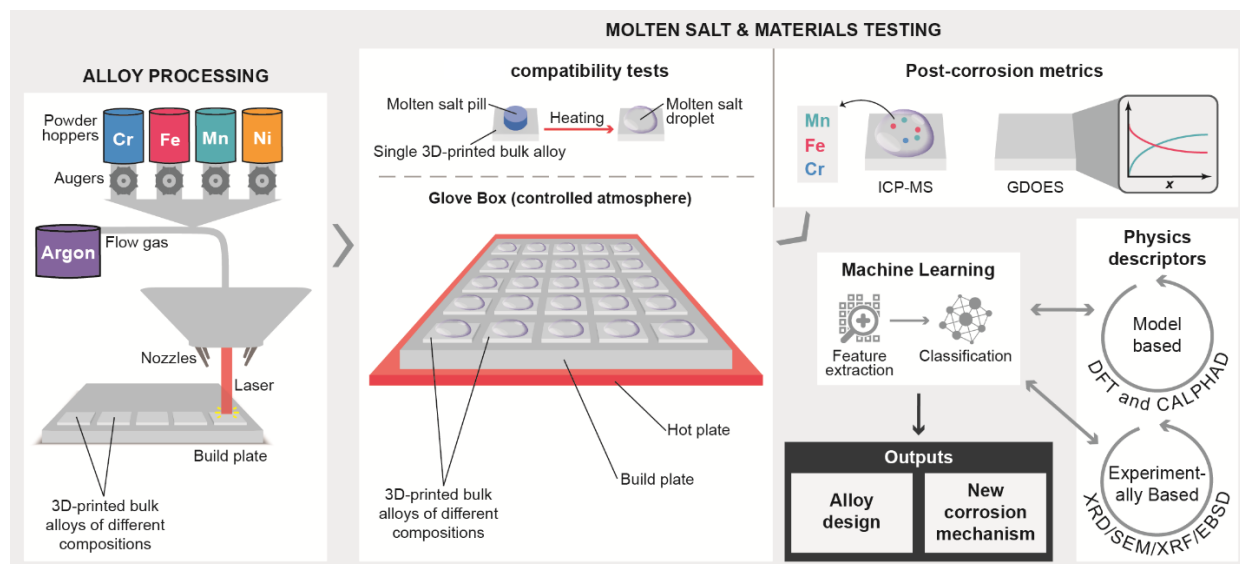

Figure 1. **Schematic of the developed HTP and automated methods for corrosion-resistant alloy development.**

## Results

### Additive manufacturing of Cr-Fe-Mn-Ni alloys

Cr-Fe-Mn-Ni alloy systems have a relatively well-known set of properties inherited from the stainless-steel family and thus represents a model alloy system to serve as a case study in the HTP platform investigation shown in Figure 1. Additive manufacturing was selected to perform in-situ alloying through the LENS process, as developed in [13]. By using this technology, Twenty-five bulk Cr-Fe-Mn-Ni alloys with dimensions of 10 x 10 x 2 mm were printed on a 316 stainless steel build plate (Figure 2(a)). The alloy compositions were targeted towards high Ni and low Cr to stabilize a single-phase FCC structure across all compositions and follow molten salt alloy design baseline requirements. To remove the dendritic compositional segregation and residual stresses, which are often observed in the as-printed alloys [14], the twenty-five printed alloys were simultaneously homogenized at 1000 °C for 24 hours in a vacuum furnace with a base pressure of $10^{-6}$ torr. A follow-up heat-treatment was also performed at 700 °C for 24 hours to obtain a stable microstructure of the printed alloys for subsequent corrosion testing. After heat treatment, the printed alloys were annealed, leveled, polished, and labelled as indicated in Figure 2(a). The compositions of the printed alloys were measured by automated energy dispersive spectroscopy (EDS) and the results are plotted in Figure 2(b). X-ray fluorescence (XRF) was also used to accelerate chemical characterization of the alloys and the results match the compositions obtained by local EDS within a few at% (see supplementary document). As expected from the targeted compositions, the final compositions are Ni-rich with very low Cr content. The detailed information regarding the composition of each printed alloy can be found in Table S-1 in the supplementary document. The equilibrium phases of the Cr-Fe-Mn-Ni system at 700 °C were simulated using CALPHAD modelling based on the PanHEA database of the Pandat software (version 2020) [15] and the results are shown in Figure 2(b), predicting that all the printed alloys are single-phase FCC.

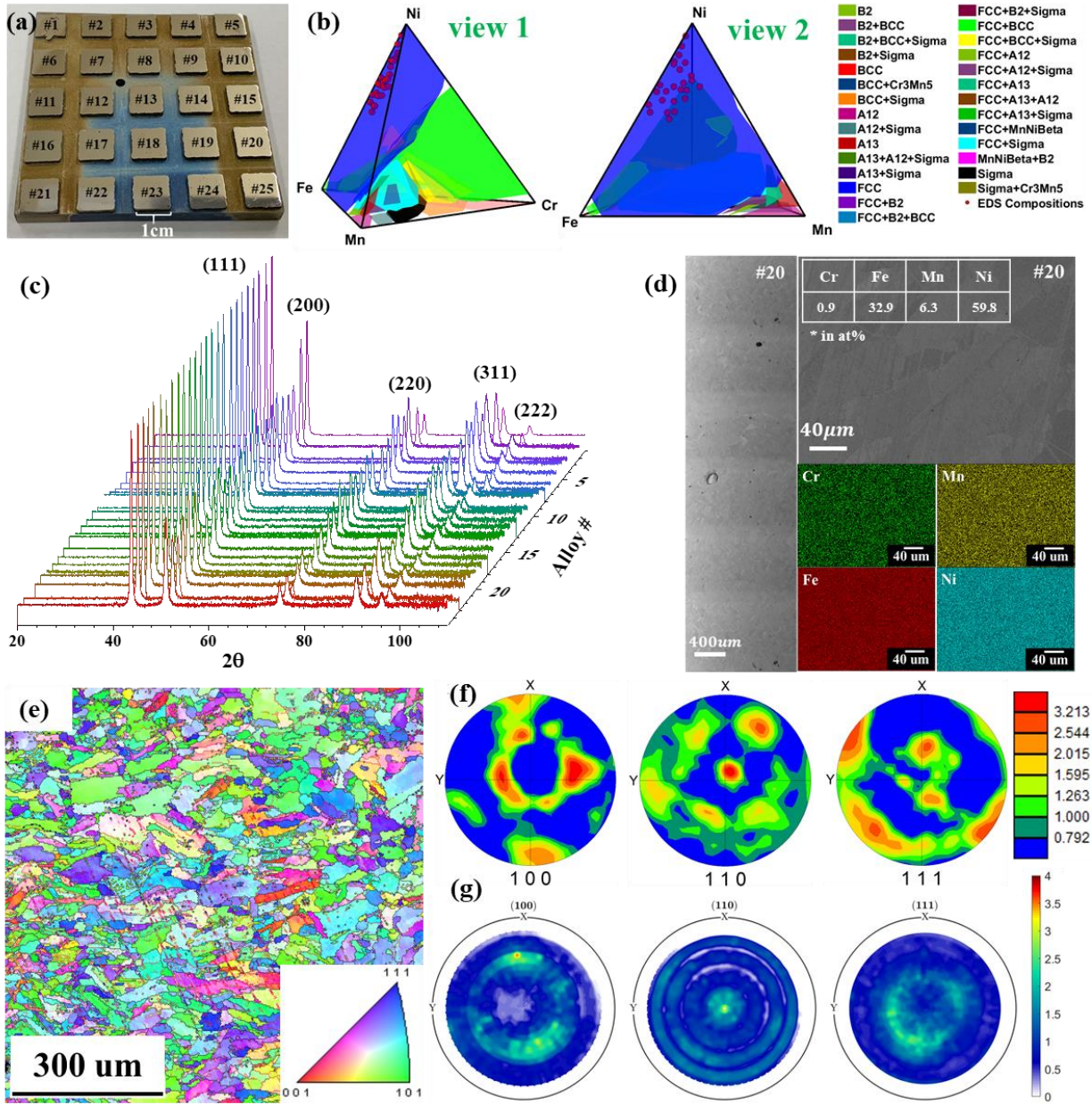

Figure 2. **Characterizations of pre-corroded Cr-Fe-Mn-Ni alloys.** (a) The twenty-five printed Cr-Fe-Mn-Ni alloys with different compositions after homogenizing, aging, and polishing. (b) Compositions of the printed twenty-five Cr-Fe-Mn-Ni alloys identified by EDS with the CALPHAD equilibrium phases of Cr-Fe-Mn-Ni alloy system. (c) XRD patterns of the twenty-five printed alloys after heat treatments (d) SEM and EDS analysis of alloy #20 (note that signal to noise ratio in the Cr map is probably quite low). (e) EBSD IPF map of the surface normal to the build direction of alloy #21 (0.9 at% Cr, 30.5 at% Fe, 14.7 at% Mn, and 53.9 at% Ni). (f) EBSD pole figures corresponding to the area shown in (e). (g) XRD pole figures of alloy #21.

Automated X-ray diffraction (XRD) was performed on each printed alloy after heat treatment and the profiles are presented in Figure 2(c). The results show that all alloys are indeed single-phase FCC, which is consistent with the CALPHAD predictions. The 2θ peak position varies slightly between each sample, which is likely due to the change in lattice parameter brought by different compositions (see Table S-2 in the supplementary document). The microstructures of all the printed alloys were examined by automated scanning electron microscopy (SEM) before corrosion except for alloy #1, which showed severe surface defects from the printing and was discarded from the study. Figure 2 (d) illustrates an example of SEM/EDS on a printed alloy surface: alloy #20 from which no defects of cracking or craters were found on the polished alloy surfaces. Spherical porosity and defects from inadequate fusion were sporadically observed as is typical in additive manufacturing technology [16]. These pores and defects could be due to the un-melted powder particles during the additive manufacturing process. Higher magnification SEM micrographs showed limited evidence of microscale printing defects. Elemental EDS mapping also revealed that the printed alloys are compositionally homogeneous. These automated characterization methods demonstrate that the LENS in-situ alloying is an ideal HTP method to manufacture bulk alloys for corrosion experiments. Additional SEM images and EDS mapping for other printed alloys are shown in Figure S-1, Figure S-2, and Figure S-3 in the supplementary document. It is worth mentioning that all heat treatments and characterizations were performed while the samples were still attached to the build plate, allowing for automation of the different processes. Electron backscatter diffraction (EBSD) was also performed on the top surface of multiple printed alloys to characterize grain morphology. As a typical example, Figure 2(e) and (f) display the EBSD scan results of alloy #21 (0.9 at% Cr, 30.5 at% Fe, 14.7 at% Mn, and 53.9 at% Ni). Figure 2(e) shows the inverse pole figure (IPF) map displaying the overall grain morphology and orientation at the surface. A distribution of small equiaxed grains and larger columnar grains was observed. These columnar grains are irregularly shaped and a majority of them have a high aspect ratio. The microstructure of the alloy exhibits a strong (110) texture as evidenced by the EBSD pole figures in Figure 2(f). To increase the throughput of the sample texture information, XRD pole figures were also performed on the same set of alloys. The XRD pole figure results on alloy #21 are shown in Figure 2(g) which matches the EBSD results quite well, showing the relatively high (110) surface texture at the center. This means that (110) is prevalently parallel to the alloy surface. All printed alloys exhibited similar grain morphologies and texture, resulting from similar in-situ alloying parameters used in the LENS manufacturing process (Figure S-4 and Figure S-5 are for additional alloys).

**HTP materials/molten salt compatibility test**

HTP molten salt corrosion experiments were carried out on each of the printed alloys, while still attached to the build plate, to study their corrosion resistance in molten salt. LiCl-KCl (44wt% LiCl-56wt% KCl, melting point: 353 °C) eutectic salt, the extraction media of pyroprocessing of spent nuclear fuel, was used for the corrosion experiment. 2 wt% $EuCl_3$ salt was added into LiCl-KCl eutectic salt to increase the overall corrosion driving force by increasing the redox potential of the salt [17], such that the corrosion effects could be observed over a relatively short exposure time. To perform the corrosion test, salt pills consisting of LiCl-KCl-2wt% $EuCl_3$ were prepared (Figure 3(a)) by a standard salt pelletizing procedure. The mass of each salt pill prepared for the corrosion test is illustrated in Figure 3(a) which shows that the pill's mass was controlled in the range of 0.368 g to 0.385 g. Due to the relatively small variations in salt pill mass, the volume of salt to alloy surface area ratio is considered constant for all the tested alloys. The solid salt pills were placed on the polished surface of each printed alloy and melted into droplets at the temperature of 500 °C by a heating plate as described in the compatibility test of Figure 1. The corrosion test was performed for 96 hours inside an inert atmospheric glovebox ($O_2$<2 ppm, $H_2O$<0.1 ppm). After the corrosion test, the melted molten salt droplets were cooled down to room temperature, solidified, and extracted.

Each printed alloy was cleaned ultrasonically after corrosion test and a series of automated material characterization analyses were performed to characterize the corrosion performance. First, SEM was used to evaluate the corrosion attack on the surface of each printed alloy. The SEM micrographs of a few representative printed alloys are shown in Figure 3(c). Differences in surface corrosion attack of the various alloys is evident. Detailed information of the surface corrosion attacks of other printed alloys are presented in Figure S-6 and Figure S-7 in the supplementary document. Automated XRD was used to identify any possible phase changes in the near surface regions of the printed alloys during the corrosion test. In Figure 3(d), small peak splitting was observed on the XRD patterns of the post-corrosion alloys. Similar phenomenon also appeared on the post-corrosion Ni-201 alloy as reported in a previous study [18]. However, in the present study, the extent and 2θ location of peak splitting were found to vary among the different printed alloys (see XRD patterns of all other post-corrosion printed alloys in Figure S-8 in the supplementary document). The peak splitting is likely a result of the change in lattice parameter at the near surface region, induced by the changes in composition of the FCC phase (depletion of elements, and injection of vacancies). Overall, no significant phase change was observed, confirming that the alloys remained single-phase FCC during the corrosion test.

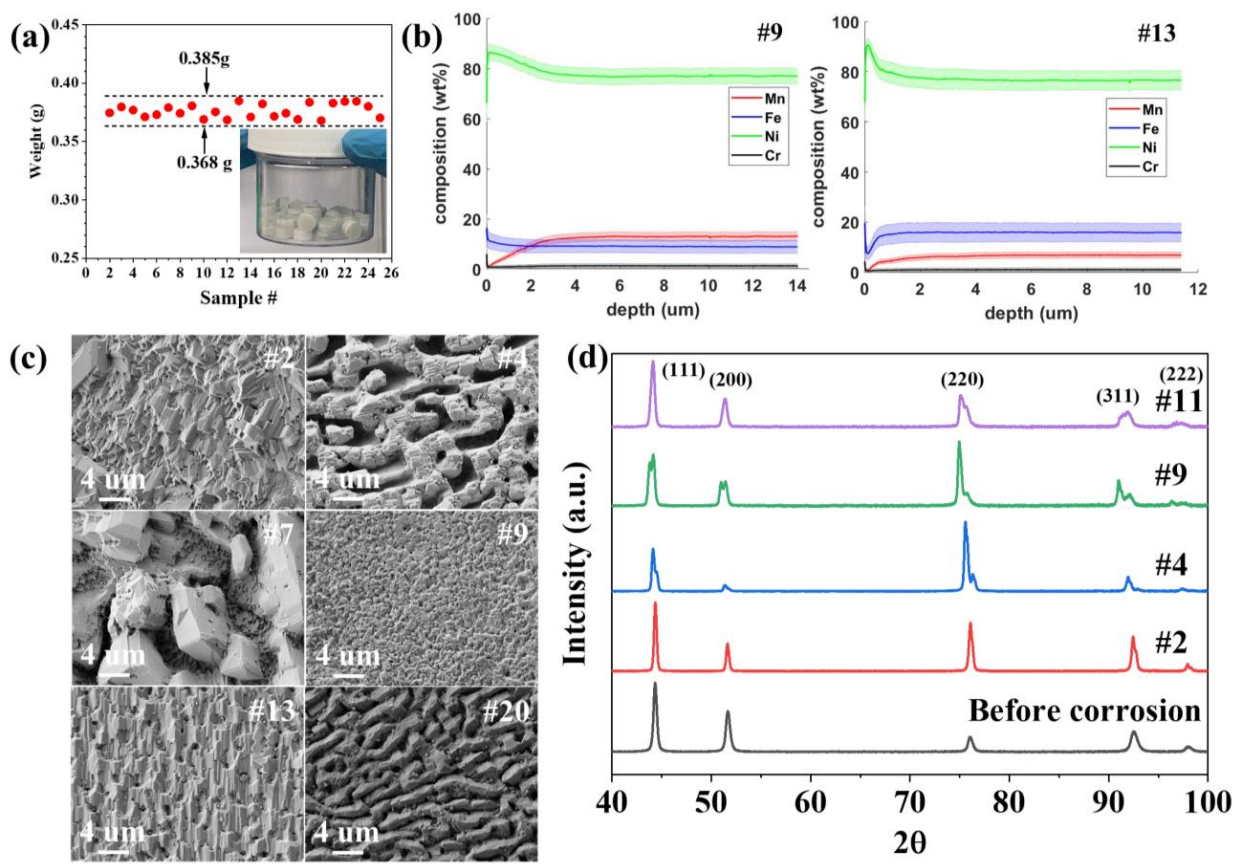

Figure 3. **HTP corrosion test of the printed alloys in LiCl-KCl-2 wt% EuCl$_3$ molten salt at 500 °C.** (a) Salt pills and their mass for corrosion test on each corresponding printed alloy. (b) GDOES analyses of two representative printed alloy after corrosion test, alloy #9 and alloy #13 (the shade area represents one standard deviation of the measured value). (c) SEM micrographs of the top surfaces of a few representative printed alloys after corrosion test: alloy #2, alloy #4, alloy #7, alloy #9, alloy #13, and alloy #20. (d) XRD patterns on the top surfaces of a few representative printed alloys after corrosion test: alloy #2, alloy #4, alloy #9, and alloy #11, the XRD pattern before corrosion used as a reference is taken from alloy #2.

To compare corrosion resistance, the corrosion attack depth was measured by automated GDOES analysis technique. Figure 3(b) gives an example of the obtained compositional profiles of Cr, Fe, Mn, and Ni as a function of depth (0 being the salt/sample interface) for the post-corrosion alloy #9 (1.3 at% Cr, 12.6 at%

Fe, 19.3 at% Mn, and 66.7 at% Ni) and alloy #13 (0.7 at% Cr, 19.2 at% Fe, 11.4 at% Mn, and 68.7 at% Ni) by GDOES (the GDOES profiles for other alloys are presented in Figure S-9 in the supplementary document). The elemental variations as a function of depth observed from the GDOES profile result from the elemental dissolution during corrosion. Based on Figure 3(b), Mn is significantly depleted after the corrosion of alloy #9 while Mn and Fe are both depleted for alloy #13. Similarly, the depleted elements for the other printed alloys were found to be Mn or/and Fe as shown in the supplementary document. The depth corresponding to the variation of the depleted element usually signifies the "corrosion attack depth" or "depletion depth". To systematically report the corrosion attack depth for each of the post-corrosion printed alloys, an algorithm was developed to analyze the data measured by GDOES (see Methods section). The depletion depths of a given element for the different printed alloys are automatically generated and the results are shown in Figure 4(a). The data in Figure 4(a) shows that alloy #2 (0.6 at% Cr, 10.4 at% Fe, 1.1 at% Mn, and 87.9 at% Ni) has the lowest depletion depth (~1.2 µm), while alloy #21 (0.9 at% Cr, 30.5 at% Fe, 14.7 at% Mn, and 53.9 at% Ni) has the highest depletion depth (~10.2 µm) for a given element. Thus, alloy #2 is the most corrosion resistant whereas alloy #21 has the lowest corrosion resistance among all the alloys tested. ICP-MS compositional analysis was also performed on the entire solidified salt droplet extracted from each printed alloy surface after corrosion test. The measured concentrations of the dissolved ions in molten salt are presented in Figure 4(b), showing that Ni and Cr barely dissolved into the salt. This is expected since Ni has a higher standard redox potential for its metal chloride formation, and the printed alloys contain very little Cr. On the other hand, significant Fe and/or Mn concentrations were observed in the post-corrosion salts. Figure 4(b) also shows that alloy #2 has the lowest amount of corrosion products dissolved in the salt while it is the highest for alloy #21. These results agree well with the depletion depth comparison obtained by GDOES. However, it should be noted that, as expected, the two performance metrics, depletion depth and overall concentration of the dissolved ions, do not necessarily follow the exact same alloy ranking order of corrosion resistance. For instance, an alloy with low Mn content could potentially lead to relatively low concentration of dissolved Mn in the salt but a significant Mn depletion depth.

To study the cross-sectional microstructure and microchemistry of the post-corrosion printed alloys, four alloys (#5, #7, #17, and #20) were cut through the center for SEM/EDS line scans and compositional maps. Figure 4(c) shows the SEM image, EDS compositional maps, and EDS line scan of alloy #20 (0.9 at% Cr, 32.9 at% Fe, 6.3 at% Mn, and 59.8 at% Ni) (similar results for the other three alloys are presented in Figure S-10 in the supplementary document). In Figure 4(c), significant subsurface voids are observed. EDS mapping shows significant Fe and Mn depletions in the near surface region. EDS line scans were also performed and the corrosion attack depths were found to be about 6.3 µm for Fe and 6.5 µm for Mn. This data is included in Figure 4(a) together with the other results obtained by GDOES for the corrosion resistance comparison of different printed alloys.

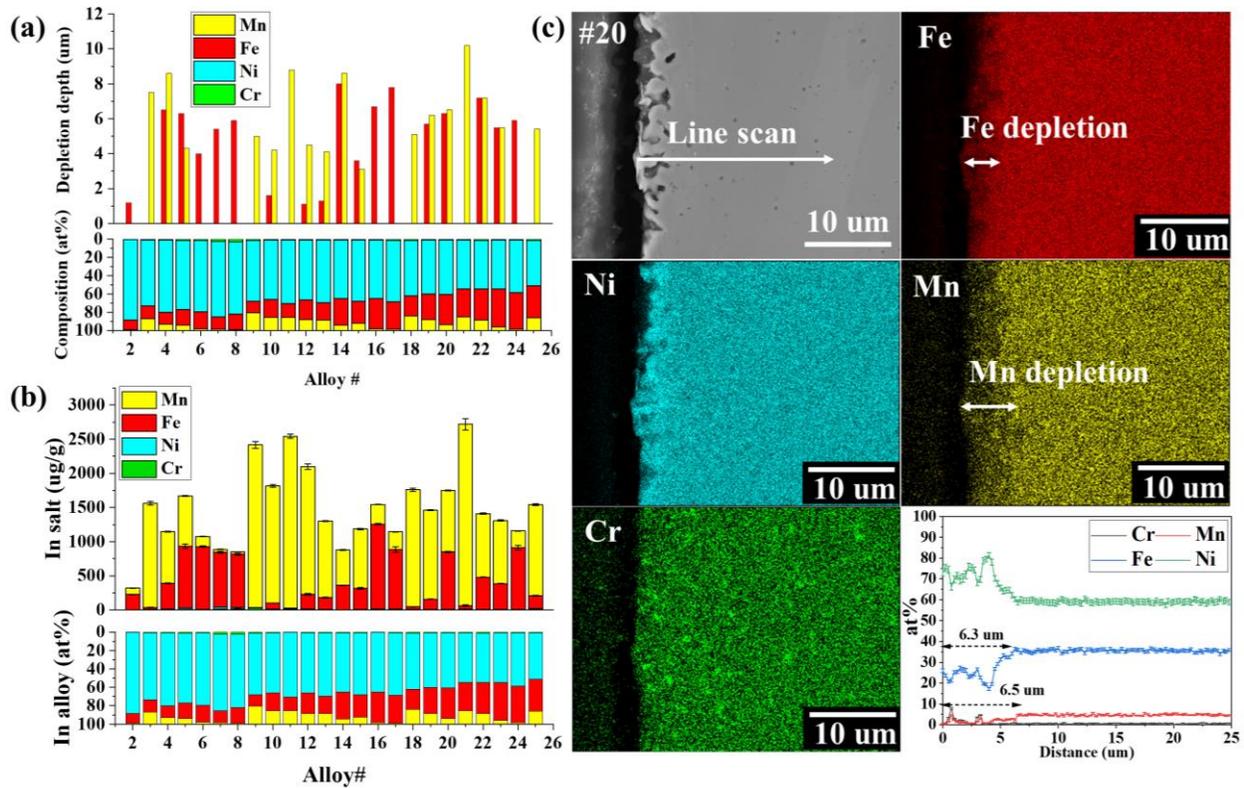

Figure 4. **Corrosion resistance identification.** (a) Corrosion attack depth of different printed alloys. (b) Concentrations of dissolved ions in molten salt during the corrosion testing for different printed alloys measured by ICP-MS. (c) SEM, EDS compositional mapping, and EDS line scan of the near surface region of alloy #20 (0.9 at% Cr, 32.9 at% Fe, 6.3 at% Mn, and 59.8 at% Ni).

**Thermodynamic/Kinetic Modelling**

Thermodynamically, work function and surface energy have been shown to be proportional and inversely proportional to a surface's stability in an oxidizing media, respectively [19]. Therefore, they could represent good indicators to rank the inherent interface stability of the printed alloys. In the present study, first-principle DFT combined with CALPHAD was used to calculate the surface energy and work function of the three common planes (111), (110), and (100) of the FCC structure of all the printed alloys and the results are shown in Figure 5(a). As can be seen from this figure, (111) appears to have lower surface energy and higher work function compared with (110) and (100) planes. This means that (111) is the most stable interface among the three planes. The ratio of surface energy to work function of the (110) plane, the most prevalent plane paralleling to the sample surface according to the pole figures in Figure 2(f) and Figure 2(g), is used to compare the corrosion resistance of different printed alloys based on the interface stability principle. Figure 5(b) shows an increasing trend between the ratio of surface energy to work function and the overall concentrations of the dissolved ions in post-corrosion salt, suggesting that the alloy corrosion resistance could be related to the interface stability.

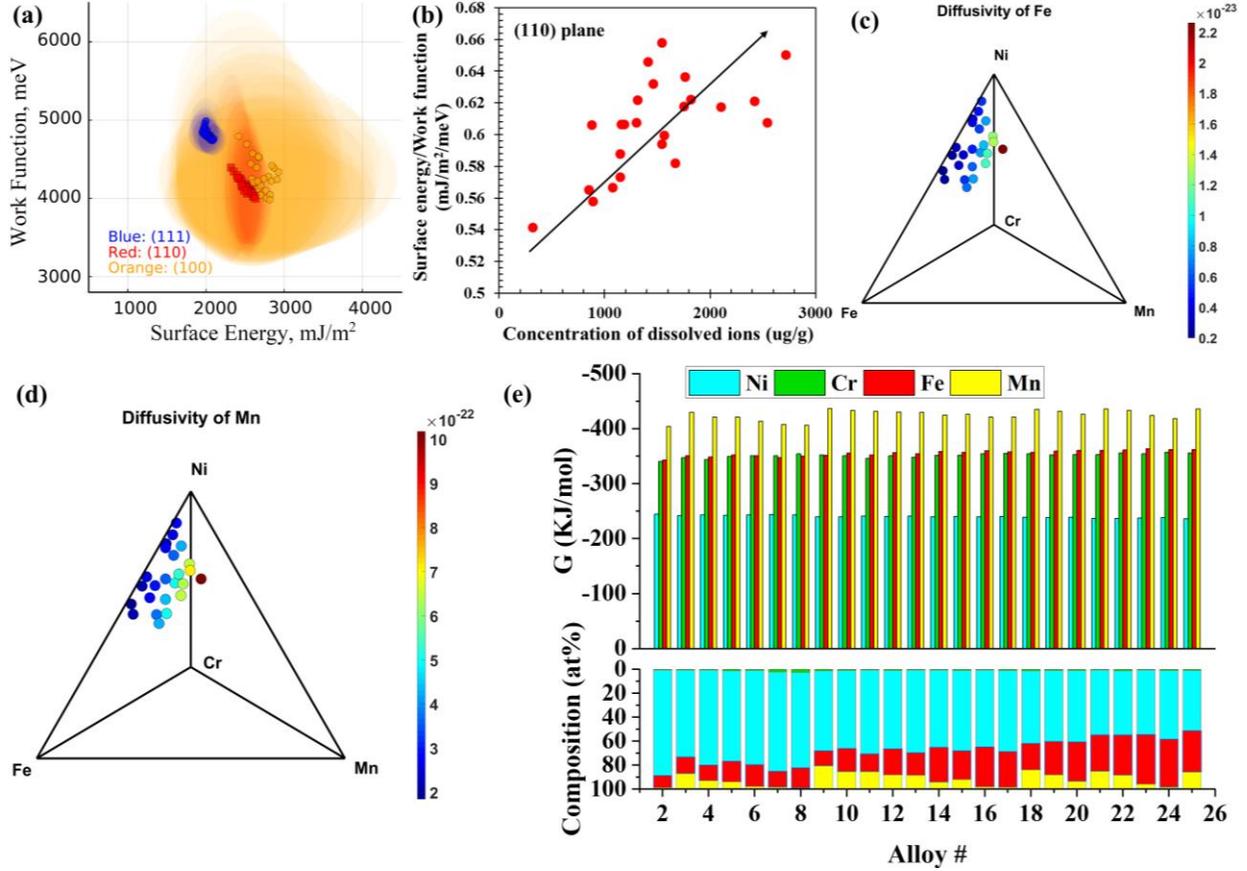

Figure 5. **Thermodynamic and kinetic simulation.** (a) Surface energy and work function of planes (111), (110), and (100) of the printed alloys (shadowed eclipses represent 95% confidence bands/uncertainties). (b) The correlation between the ratio of surface energy to work function and the overall concentrations of all the dissolved ions in post-corrosion salt (arrow line is added to guide the eye). c) Tracer diffusivities of Fe in different printed alloys (in the unit of m$^2$/s). (d) Tracer diffusivities of Mn in different printed alloys (in the unit of m$^2$/s). (e) Gibbs free energy of CrCl$_2$/Cr, FeCl$_2$/Fe, MnCl$_2$/Mn, and NiCl$_2$/Ni for different printed alloys.

Other parameters expected to influence the corrosion resistance of elements in molten salts were also derived in this study. Based on the findings in the corrosion experiment study, Mn and Fe are the main depleted elements of the printed alloys during the corrosion. The Mn and Fe dissolution rates are very likely dependent on their diffusivities within the alloy matrix. Hence, the tracer diffusivities of Mn and Fe in the different printed alloys were calculated by the kinetics module in Pandat software [15] using a mean field extrapolation approach on the databases reported in a previous study [20]. Figure 5(c) and Figure 5(d) show that the calculated tracer diffusivity of Mn is higher than that of Fe by about two orders of magnitude across all the printed alloys. Note that even if this method does not consider a specific diffusion mechanism, it is still consistent with the study [21] in which Fe is reported to have a larger energy barrier for vacancy migration compared to Mn in equimolar Cr-Fe-Mn-Ni alloy. The Gibbs free energy of metal chloride formation (CrCl$_2$/Cr, FeCl$_2$/Fe, MnCl$_2$/Mn, and NiCl$_2$/Ni) were also calculated for different printed alloys using Equation (1) as done in [22]:

$$G_{MCl_2/M} = G^o_{MCl_2/M} + RT \ln\left(\frac{\alpha_{MCl_2}}{\alpha_M}\right) \tag{1}$$

where $M$ represents Cr, Fe, Mn, and Ni, $MCl_2$ is the metal chloride formed in molten salt during the corrosion, $G^O_{MCl_2/M}$ is the standard Gibbs free energy of formation, which can be compiled from thermodynamic database, such as HSC Chemistry 6.0 ® [23], $R$ is the gas constant, $T$ is the temperature (773 K in this study), $\alpha_{MCl_2}$ is the activity of metal chloride in molten salt which is assumed to be $10^{-6}$ [22], and $\alpha_M$ is the activity of $M$ in the printed alloy calculated using the PanHEA database of Pandat software (version 2020) [15]. As can be seen from Figure 5(e), Mn has the lowest Gibbs free energy followed by Fe, Cr, and Ni. This means Mn is the element most thermodynamically reactive to form a dissolved metal chloride while Ni is the least reactive element.

**Discussion**

The results include the exposure of twenty-four different single-phase FCC alloys from the Cr-Fe-Mn-Ni system to chloride salts at 500°C for 96 hours. The results mostly show Fe and/or Mn dissolutions, and four parameters are used to define the alloy corrosion resistance, namely the Fe and Mn depletion depths and the Fe and Mn dissolved concentrations in the salt. Alloy #2, which has the highest Ni content, shows the best corrosion resistance. This is expected since Ni is the most noble element in the alloy system of interest. However, ranking other alloys' corrosion resistances based on their compositions or physical properties is more difficult. To better assess the physical parameters affecting the corrosion resistance of the printed alloys, a ML algorithm based on the Random Forest Regression (RFR) model [24] was developed. The objective of the ML-based approach is to search for the most influential physical parameters in determining the concentrations of the dissolved ions in the post-corrosion salt. In the developed model, the physical parameters selected as input features to train the active learning model include: (i) element composition, (ii) activity in the single FCC phase, (iii) metal chloride Gibbs free energy of formation, (iv) self-diffusivity in the alloy, (v) (110) plane surface energy and (vi) (110) plane work function. Figure 6(a) and (b) show the RFR 5-fold Cross-Validation (CV) predicted concentration of Mn and Fe in post-corrosion salt with respect to the measured values by ICP-MS from which a high consistence can be observed. The high correlation between CV predictions of dissolved Fe and Mn in post-corrosion salt and the experimental values measured by ICP-MS (Pearson's R value [25] is 0.797 for Fe and 0.891 for Mn, respectively) also demonstrates the robustness of the RFR model and the reliability on the extraction of the importance rankings of input features. The extracted ranking of the input features for contributing to the depletions of Mn and Fe by importance scores are shown in Figure 6(c) and (d). It was found that the physical parameters of Mn in the system, such as the composition of Mn in the alloy and Gibbs free energy of $MnCl_2/Mn$, consistently ranks at the top in the importance score ranking for the predictions of Mn and Fe dissolved in the salt. This indicates that the behavior of Mn in the alloy plays a crucial role on the dissolutions of both Mn and Fe from the alloy into the molten salt. The effect of Mn related features on Fe dissolution indicate a possible sacrificial mechanism of Mn to protect Fe. Based on this finding, the Fe/Mn composition ratio was added as an input feature into the RFR model to study its potential influence on the depletion of Fe. The extracted input feature ranking in Figure 6(e) and (f) shows Fe/Mn composition ratio contributes the greatest to the depletion of Fe.

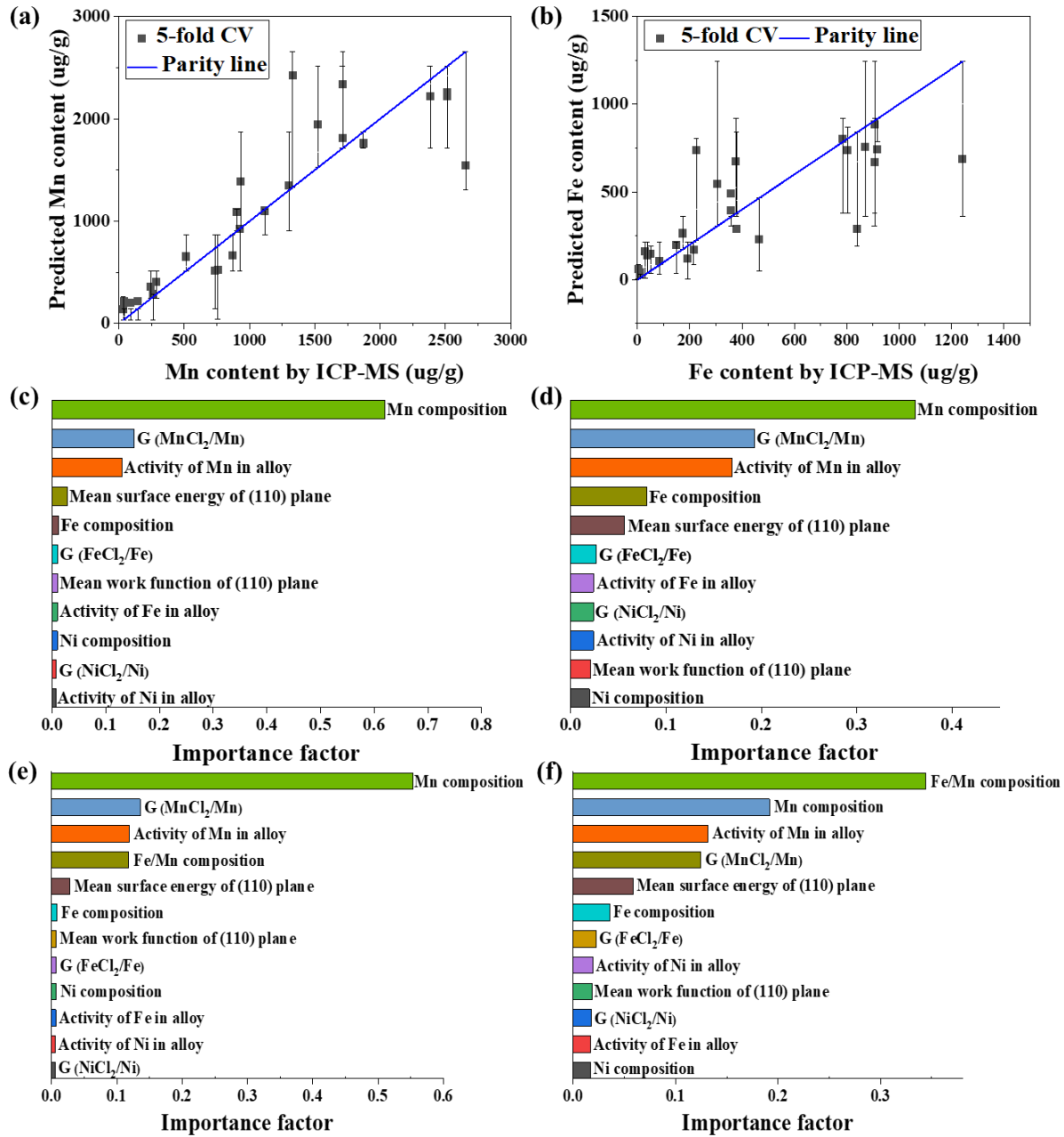

Figure 6. **ML for corrosion resistance prediction.** Predicted vs. experimental values of (a) Mn and (b) Fe concentrations in post-corrosion salt (error bar represents the 95% confidence interval of the prediction values). Input feature relative importance ranking order for contributing to the corrosions of (c) Mn and (d) Fe in molten salt without the adding input feature of Fe/Mn composition ratio. Input feature relative importance ranking order for contributing to the corrosions of (e) Mn and (f) Fe in molten salt with the added input feature of Fe/Mn composition ratio.

Based on the results presented by the RFR model, the two most important factors that affect the corrosion of Mn from the printed alloys, Mn composition and Gibbs free energy of $MnCl_2/Mn$, are discussed in greater detail. In Figure 7(a), the dissolved concentration of Mn is plotted as a function of the Mn composition in the alloy. A clear and expected trend is observed where a higher Mn content in the alloy results in more Mn dissolved in the salt. This means that the dissolution of Mn from the printed alloys gets worse with the increase of Mn content in the alloy. From the perspective of thermodynamics, the $MnCl_2$/Mn couple (in different printed alloys) with a lower Gibbs free energy of formation is more likely to have Mn be depleted by molten salt and needs more Mn to dissolve into molten salt such that an equilibrium can be reached between Mn activities in the salt and in the alloy. As a result, an inverse correlation between the dissolved Mn content in molten salt and the Gibbs free energy of $MnCl_2$/Mn, calculated by Equation (1), is observed as shown in Figure 7(b). The dissolution rate should also be dependent on the elemental diffusion from the bulk to the alloy surface, namely, the diffusivity. The Mn content dissolved in the salt is plotted as a function of its tracer diffusivity in the alloy in Figure 7(c) and a positive correlation is found. However, the tracer diffusivity was of no importance to Mn dissolution in the ML relative importance ranking order. This means that while the correlation exists (Figure 7(c)), its importance within the 24 alloys examined in this study is quite weak relative to the other physical parameters, such as the Mn composition and chloride Gibbs free energy of formation.

Fe is the other element that is depleted in the alloy due to corrosion. It has been observed that, unlike Mn, no correlation was observed between the Fe physical parameters and the depletion of Fe from the alloy (see Figure S-11 in supplementary document). On the other hand, the Mn related physical parameters, such as Fe/Mn ratio and Mn composition in the alloy, rank at the top in the importance ranking for the depletion of Fe. Thus, the ML indicates that the depletion of Fe will be greatly affected by the behavior of Mn in the printed alloy. A possible mechanism for this interdependency is detailed below based on the concept of "sacrificial protection". During the corrosion process, Mn and Fe atoms at the surface will be first depleted and dissolve into molten salt since the salt redox potential, fixed by the europium redox couple, is higher than that of $FeCl_2$ and $MnCl_2$ [17,26]. Following the dissolution, vacancies will be injected in the alloy subsurface (see Figure 4(c)). From our kinetics results, the diffusivity of Mn in the alloy is two orders of magnitude higher than that of Fe (Figure 5(c) and (d)), such that it is much more likely that, if a Mn atom is adjacent to a vacancy, the vacancy will be filled by the Mn atom rather than by an adjacent Fe atom. Consequently, the vacancy flux induced by the atomic dissolution at the surface will result primarily in a Mn flux to the alloy surface by inverse Kirkendall effect [27]. In this mechanism, the Mn "sacrifices" itself and protects the Fe from being depleted, and the Mn composition (i.e. the probability for a Mn atom to be adjacent to a vacancy) ranks the highest in contributing to the elemental dissolution. This is also verified by the correlation between the Fe/Mn ratio in the alloy and the dissolved Fe content as plotted in Figure 7(d), implying that the depletion of Fe can be mitigated by increasing the Mn/Fe ratio. The output of the sacrificial protection mechanism is quite analogous to the well-known sacrificial anode mechanism observed in room temperature aqueous corrosion [28]. However, the proposed mechanism is quite unique since it is based on in-situ alloying sacrificial mechanism in a homogeneous alloy, rather than a physical separation of the cathode and anode. To the authors knowledge, this in-situ sacrificial alloying during high-temperature corrosion has not been reported in the past. This unique mechanism is enabled by the active alloying element dissolution in molten salts and the thermally activated diffusion resulting from the high temperature of the environment. While this approach does not lead to a corrosion resistant alloy (i.e. an alloy that would not dissolve any element in aggressive molten salt conditions), it opens new avenue in alloy design for molten salt applications. For instance, an alloying element that is necessary for mechanical properties, but is, on the other hand, electrochemically active, such as Fe, can be protected from dissolution in molten salts by the addition of another electrochemically active and fast diffusing element, such as Mn.

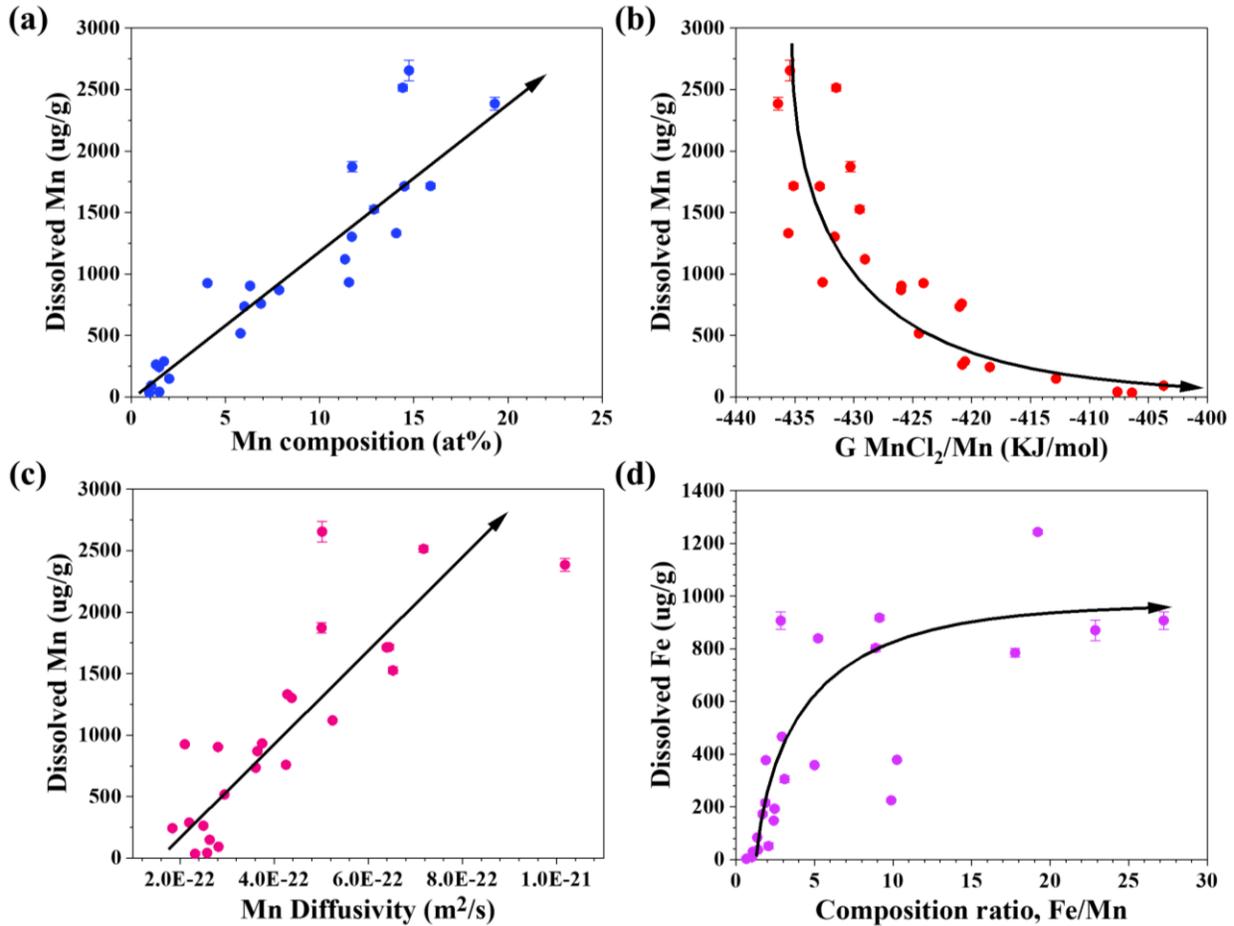

Figure 7. Binary correlations of physical parameters with the dissolutions of Mn and Fe in molten salt (arrow line is added to guide the eye).

This study developed and utilized a series of innovative HTP experimental and modeling methods to accelerate the discovery of molten salt corrosion-resistant materials. Comparing with the traditional alloy processing, such as arc melting, and typical corrosion testing in which only one alloy can be examined at a time, the in-situ alloying [13] and the HTP corrosion testing in this study has the potential to save orders of magnitude of time. Advanced material characterization methods, such as GDOES, without the need to cut and polish the post-corrosion samples as done in traditional approaches can significantly reduce the characterization time as well. Considering that the in-situ alloying method used in this study [13] was assessed to increase the sample processing time by about two orders of magnitude, the total time savings in this study from sample processing to post-corrosion data analytics is expected to be at least three orders of magnitude. The methodology could be adopted for other alloy systems to accelerate material development. Using this integrated approach for the first time, a sacrificial mechanism was found in the Cr-Fe-Mn-Ni system, providing new insights on alloy design of structural materials for molten salt applications. For instance, increasing the composition ratio of the unstable element to the less unstable element (Mn/Fe in this study), pending that the diffusivity of the unstable element is relatively higher, can prevent specific elements from dissolving in the salt (Fe in this case). The developed HTP platform, demonstrated in this study, could play a very important role in the rapid development and down-selection of corrosion-resistant alloys to support the deployment of molten salt technologies.

**Methods**

**Additive manufacturing.** In-situ alloying via directed-energy deposition (DED) based additive manufacturing was utilized to manufacture the alloys in a high-throughput manner. The fabrication was performed in an Optomec LENS MR-7 system using gas-atomized elemental powders of Cr, Fe, Mn, and Ni with a size distribution of ~45-150 µm. Moorehead et al. [13] demonstrated the technique of in-situ alloying via DED. This process was further modified and refined in this work for the Cr-Fe-Mn-Ni system.

In Optomec LENS MR-7 system, each powder hopper was filled with one of the elemental powders: Cr, Fe, Mn, or Ni. Flowing Ar gas carried the metal powders from the hoppers into the path of the laser beam. The quantity of the powder to be blown was controlled by the RPM of the augers on the individual powder hoppers. Twenty-five alloy samples, of nominal dimensions $10 \times 10 \times 2$ mm, were printed on 316L stainless steel build plates in 5×5 arrays with a spacing of 6.35 mm between samples. Each sample was comprised of five print layers, with two remelting passes after each print layer. The hatch spacing was 0.381 mm for the print layers and 0.19 mm for the remelting layers, with a 90° rotation of the hatch pattern between each subsequent print/remelt layer. Detailed description of the process development can be found in [13].

**Material preparations and corrosion experiments.** The LiCl-KCl eutectic salt used in this study was prepared with 44 wt% of anhydrous LiCl salt powder (≥ 99% purity, Sigma Aldrich) and 56 wt% of anhydrous KCl salt powder (≥ 99% purity, Sigma Aldrich). 2 wt% of anhydrous $EuCl_3$ (99.99% purity, Sigma Aldrich) was added into LiCl-KCl eutectic salt to increase the overall corrosion rate by increasing the redox potential of the salt. The LiCl, KCl, and $EuCl_3$ salt powders were mixed homogenously according to their corresponding compositions. The trace impurities of the salt mixture were identified by ICP-MS analysis in which the ten most prominent impurity elements and their concentrations were 16.5 ppm Na, 7.5 ppm Mg, 1.1 ppm Al, 3.8 ppm P, 11.2 ppm S, 32.3 ppm Ca, 0.2 ppm Cr, 1.0 ppm Fe, 0.02 ppm Mn, 0.1 ppm Ni. For salt pelletizing, approximately 0.375 g salt mixture was weighed, transferred into a custom-fabricated tungsten carbide die, and compacted through a Pike Pixie Manual Hydraulic Press at a constant load of 2.5 tons for 2 minutes. The completed salt pill was then punched out by the manual hydraulic press and each salt pill was confirmed to be compact enough as a precaution that ensures minimal mass change of the pill in transit. The entire salt pill preparation process was performed inside an argon atmosphere glovebox ($O_2$<2 ppm, $H_2O$<0.1 ppm). Before corrosion experiment, the printed alloys were ground with SiC abrasive papers of different grit sizes up to 1200 grit following which they were polished on the polishing pads by 3 um, 1 um diamond suspensions, and 0.04 um colloidal silica suspension. Then the polished alloys were ultrasonically cleaned with deionized water, ethanol and acetone.

For the corrosion experiment, the pelletized salt pill was placed on the center of the top surface of each printed alloy. Then the salt pill was melted to form a droplet at 500 °C by a heating plate inside the glovebox filled with argon gas. Due to surface tension, the molten salt droplet stayed on the surface of the printed alloy without spillage. The corrosion experiment was performed for 96 hours after which the heating plate was shut down and the entire solidified salt droplet was removed from each of the printed alloys for further analysis. A low speed saw was used to cut the post-corrosion samples of which material characterizations were performed on the cross sections. The cross sections were then polished using the same procedure as done for the printed alloys before the corrosion experiment.

**Material characterizations.** SEM coupled with EDS and EBSD was performed at the Wisconsin Center for Nanoscale Technologies to characterize the samples before and after corrosion experiment. EBSD was acquired at a magnification of 100 times and ~14mm working distance using a step size of 5 um. Before analysis, the acquired EBSD data was cleaned by eliminating grains with diameters less than 5 µm and misorientations less than 0.5°. XRD used in this study was conducted by Bruker D8 Discover diffractometer with a Cu-K$_\alpha$ micro X-ray source. XRD pole figure was normalized by an internal Bruker software and the normalized data were exported and plotted using MTEX software after background and defocusing correction. There are still some artifacts in the XRD pole figure that could be a result from the florescence and/or a sector of the detector that has lower intensity. GDOES was used to provide information regarding the concentration profile of each element along the depth of the analyzed sample through which the depletion depth could be identified without the need to cut and polish the post-corrosion samples as done in traditional material characterization approaches. The GDOES was carried out by Horiba GD-profiler 2™ using a pressure of 550 Pa and a power of 40 W for the plasma generation. The calibration of GDOES was performed using sputtering rate correction with certified materials with known compositions which covered the concentration ranges of 0~100 wt% for Ni and Fe while 0~30 wt% for Cr and Mn, respectively. The standard deviation of the measured data was derived based on the uncertainty on the concentrations of the standards, sputtering rate of the standards, the detection limit, and the measured intensity of the standards using classical error propagation formulas. The measured GDOES data was analyzed by a developed algorithm as follows to obtain the depletion depth:

1) determining an average value $\bar{C}$ and standard deviation σ of the bulk composition data of the depleted element as displayed at the right side of the GDOES profile window where the composition curves are stable;

2) searching the composition data $C_d$ from the sample surface (depth is 0) verifying $C_d = \bar{C} - 3\sigma$;

3) $C_d$ is defined as the corrosion attack depth (depletion depth).

Except alloy #5, alloy #7, alloy #17, and alloy #20 for which the depletion depth is measured by EDS line scan, all other alloys' depletion depths were obtained from GDOES analysis. The depletion depths measured by EDS was used to validate the GDOES results.

Dissolved cations in post-corrosion salt were quantified by digesting the whole solidified salt droplet after corrosion experiment into deionized water and then analyzing it using ICP-MS at the Wisconsin State Laboratory of Hygiene. One benefit of using salt pills in the corrosion experiments is that the entire pill can be dissolved for ICP-MS analysis, ensuring the measured analyte concentration in the sample matrix represents the averaged composition of the entire salt volume participating in the corrosion process.

**DFT and CALPHAD Modelling.** Surface energies and work functions of various FCC alloy compositions were calculated by DFT, implemented by the Vienna Atomistic Simulation Package (VASP) [29]. The considered compositions ranged from unary to quaternary with a composition step of 25 at% for binary and ternary and 12.5% for quaternary. Random alloy structures were modeled by 32-atom special quasi-random structures (SQS) [30] using the ATAT package [31] while surface structures were generated from the SQSs using the Pymatgen package [32]. Generalized Gradient Approximation (PBE) projector augmented wave pseudopotential [33] was used to describe the exchange-correlation contribution and collinear spin polarization was accounted. Integration in the reciprocal space were obtained over a gamma-centered Monkhorst-Pack grid with N$_{kpoint}$ ≈ 3000/N$_{atom}$ within the first Brillouin zone [34]. Cut-off energy was set to the 1.3 times of the highest constitutional ENMAX. The electronic and ionic convergence criteria were $10^{-6}/10^{-5}$ eV respectively for bulk calculations while $10^{-5}/10^{-4}$ eV respectively for surface calculations. The bulk structures' atomic positions, volumes, and shapes were fully relaxed while only atomic positions were considered for the relaxations of surface structures. These relaxations were achieved by Hermite-Gauss smearing method of Methfessel and Paxton of order 1, with a smearing parameter of 0.01 eV [35]. At the end, Final static calculations using the tetrahedron smearing method with Blöchl corrections [36] were conducted for all calculations to improve the accuracy.

The calculated surface energies and work functions of the considered FCC compositions were then used to assess CALPHAD sub-regular solution model which allows the interpolation of data at arbitrary compositions within the quaternaries. Thus, the surface energy and work function of the printed alloy compositions could be derived based on the fitted CALPHAD model. The CALPHAD sub-regular solution model can be expressed by:

$$P = \sum_i x_i + \sum_i \sum_{j \neq i} a_{ij} x_i x_j + \sum_i \sum_{j \neq i} \sum_{k \neq i,j} a_{ijk} x_i x_j x_k \tag{2}$$

where, $P$ is either surface energy or work function, $x_{i,j,k}$ is the composition of Cr, Fe, Mn, or Ni, $a_{ij}$ and $a_{ijk}$ were model parameters that were derived using the calculated surface energies and work functions. To account for the CALPHAD model's uncertainty, quantification approach based on Bayesian statistics was also adopted and the details of which can be found in literature [37].

**Machine Learning.** A vector of 15 parameters including compositions, Fe/Mn composition ratio, Gibbs free energy, self-diffusivity, activity of Mn, Fe, and Ni, mean surface energy and work function parametrizing physical properties is associated to each alloy tested ("physical descriptors/features" of the alloy). These physical descriptors were curated by considering physical parameters of an alloy which are theoretically expected to directly or indirectly influence corrosion behavior of the alloy regardless of whether their relationships to target variables are explicitly known. Upon completion of the corrosion experiment, the corrosion performance of the alloy is parametrized by two "target variables": the concentrations of corroded Fe and Mn into the post-corrosion salt. A sample set containing 28 alloys (4 alloys are from repeat test), each with 15 descriptors and 2 target variables was used to train a Random Forest Regressor (RFR) model using the Scikitlearn package [24]. The RFR model was hyperparameter-tuned to optimize the number of estimators (1000), and maximum tree depth (none). The target variables were predicted with this RFR model using 5-fold Cross-Validation (CV) to safeguard against overfitting [38]. The ranking of features by importance score was obtained from the RFR model, indicating the relative significance of variables that influence the corrosion outcome measured by target variable value, with the most influential variable ranked at the top. This ranking order was found to be reproducible with a variation in random seed. The RFR model was retrained with a series of subsets of the original feature set and the relative feature rankings were found to be reproducible as well. This indicates robustness of the model and reliability of the ranking order.

**Acknowledgements**

This study is funded by the Advanced Research Projects Agency-Energy (ARPA-E) award number DE-AR0001050. The authors wish to thank Cody Falconer and Louis Bailly-Salins on the help of experimental setup and GDOES instructions.

**Author contributions**

Y.W. constructed the experimental facility with the assistance from B.G.. Y.W. performed corrosion experiments, data analysis, and sample preparations. Y.W., P.N., R.D., and, N. H. conducted the material characterizations. P.N. and M.M. performed additive manufacturing. B.G. and J.H. contributed the machine learning. Y.W. and T.D. performed the thermodynamic/kinetic modelling. Y.W. and A.C. drafted the manuscript. A.C., K.S. D.T., and S.C. conceived of the original project and oversaw its execution, providing regular guidance.